\documentclass[prl,amsmath,amssymb,twocolumn,showpacs,floatfix,reprint]{revtex4-1}

\usepackage[utf8]{inputenc} 
\usepackage{graphicx}
\usepackage{dcolumn}
\usepackage{bm}
\usepackage{units}
\usepackage{times}
\usepackage{miller}

\begin{document}

\title{The Equivalence Between Unit-Cell Twinning and Tiling in Icosahedral Quasicrystals}

\author{A. Prodan}

\author{R. Dušić Hren}

\author{M. van Midden}

\author{H. J. P. van Midden}

\author{E. Zupanič}
\affiliation{Jožef Stefan Institute, Jamova 39, SI-1000 Ljubljana, Slovenia}

\date{\today}

\begin{abstract}
It is shown that tiling in icosahedral quasicrystals can also be properly described by cyclic twinning at the unit cell level. The twinning operation is applied on the primitive prolate golden rhombohedra, which can be considered a result of a distorted face-centered cubic parent structure. The shape of the rhombohedra is determined by an exact space filling, resembling the forbidden five-fold rotational symmetry. Stacking of clusters, formed around multiply twinned rhombic hexecontahedra, keeps the rhombohedra of adjacent clusters in discrete relationships. Thus periodicities, interrelated as members of a Fibonacci series, are formed. The intergrown twins form no obvious twin boundaries and fill the space in combination with the oblate golden rhombohedra, formed between clusters in contact. Simulated diffraction patterns of the multiply twinned rhombohedra and the Fourier transform of an extended model structure are in full accord with the experimental diffraction patterns and can be indexed by means of  three-dimensional crystallography.
\end{abstract}


\maketitle

\textit{Introduction.}--Ever since quasicrystals (QCs) were first reported \cite{1} they attracted great interest, because they apparently contradicted some basic concepts of crystallography \cite{2,3,3a}. Contrary to fully disordered solids and perfectly grown single crystals, characterized by their rotational and translational symmetries, QCs with their forbidden five-fold rotational symmetry and with the apparently lost translational order represented something in between the two categories. However, some of their properties contradict this distinction. Their shapes can be well developed and the corresponding diffraction patterns (DPs) show exceptionally sharp reflections, without any diffuse scattering, characteristic of short-range order, modulation, or any other deviation from an ideal crystalline structure.

Pauling was convinced that none of the existing crystallographic rules was violated in the newly discovered materials \cite{4,5,6,7,8,9}. He believed these crystals were composed of twinned cubic domains with huge unit cells, whose basic building elements were composed of one Mn atom linked to twelve Al atoms. Although the existing experiments seemingly supported his model, he after all run into problems. Another major problem with Pauling’s approach was, that no twin boundaries were ever detected in QCs \cite{16}. 

Contrary to Pauling, a number of researchers \cite{10,11,12,13,14,15,15a} considered QCs an exception to the known solid state structures, which required a novel approach. Their explanation was based on the so-called Amman tiling \cite{17}, the three-dimensional equivalent of the two-dimensional Penrose tiling. Likewise to two Penrose rhombic tiles filling a plane, their three-dimensional equivalents, the prolate and the oblate golden rhombohedra, will fill the space and form the QC structure.

It is shown in the present work that tiling in the icosahedral QC structure can also be properly explained by cyclic unit cell twinning \cite{18a,18b}, applied on primitive golden rhombohedra, forming thus intergrown twins without explicit twin-boundaries.

\textit{The stuctural units and their stacking.}--First, the structure of the icosahedral star polyhedra and the corresponding DPs were considered. Twinned prolate golden rhombohedra were combined into a rhombic hexecontahedron (RH) \cite{28} with the corresponding DPs constructed by overlapping individual contributions. No contributions from additional long-range ordering, formed across clusters surrounding the RHs, were considered at this stage. 

The prolate $(a=\unit[0.435]{nm}, \alpha_{p}=\unit[63.43]{^{\circ}})$ and the oblate $(a=\unit[0.435]{nm}, \alpha_{p}=\unit[116.57]{^{\circ}})$ golden rhombohedra with $\alpha_{p} + \alpha_{o} = 180^{\circ}$ \cite{20,21} were obtained by deforming a parent face-centered cubic structure with a random occupation of both constituent atoms, in accord with the stoichiometry of the compound. The unit cell edge was chosen to fit the experimental DPs of MnAl$_{6}$, a representative of the icosahedral QCs. The rhombohedral angle was determined to fit the five-fold cyclic twinning with a rotational angle of $(360/5)^{\circ} = 72^{\circ}$  around the rhombohedral $\hkl<1 0 0>$ axes and with the twin-planes corresponding to the rhombohedral $\hkl{1 0 0}$ planes. The diagonals along the three-fold axes make $d_{p} = \unit[1.037]{nm}$ for the prolate and $d_{o} = \unit[0.245]{nm}$ for the oblate golden rhombohedra. In case of  MnAl$_{6}$ and for the given parameters both constituent atoms are obviously too large (the atomic radii of Mn and Al are $\unit[0.135]{nm}$ and $\unit[0.143]{nm}$, respectively) to be accommodated along the short diagonals of the oblate units. Consequently, the positions connecting the diagonals of the oblate rhombohedra can be only partly occupied. The shapes of both types of golden rhombohedra are determined by the shapes of their rhombic faces, whose diagonals are determined by the golden ratio $\varphi = (1+\sqrt{5})/2 \approx 1.61803$. Twenty prolate units are needed to construct a RH, whose origin represents the center of the cluster formed around it. The remaining empty spaces between adjacent clusters of various sizes, which cannot be filled by the prolate rhombohedra, represent the oblate rhombohedral interstices of altogether thirty possible orientations.

The DPs of the twinned prolate and oblate rhombohedra were constructed under two assumptions. First, in accord with the experimental DPs, where the first-order low-index reflections appear much stronger in comparison with the weak second-order ones, only reflections with Miller indices $\overline{1}\overline{1}\overline{1} \le hkl \le 111$ were included. Second, to exclude reflections from higher Laue zones the lengths of the "spikes" were kept below $\Delta S \leq \unit[0.18]{nm^{-1}}$. All rhombohedra, belonging to the same RH, were interrelated by rotating the starting rhombohedron into all possible twinned positions. 

\begin{figure}[ht!]
\includegraphics[width=7.5cm]{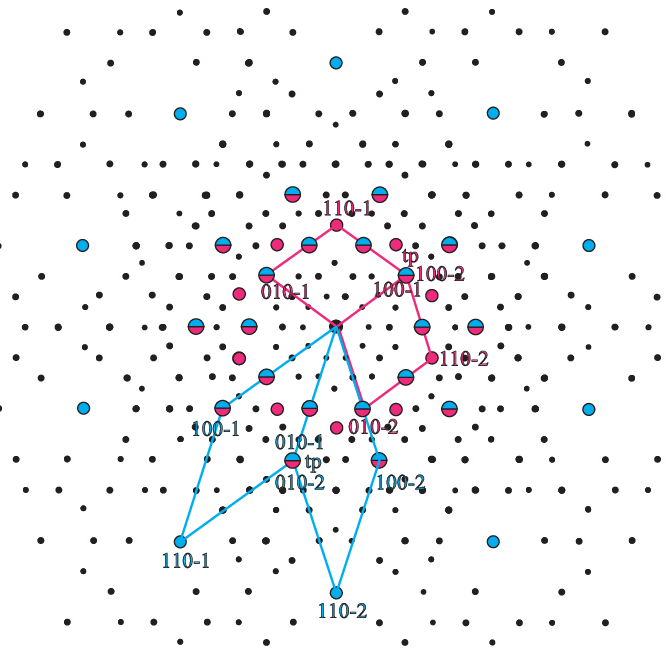}
\caption{\label{fig:fig1} The simulated diffraction pattern along one of the rhombic hexecontahedron five-fold axes, obtained by overlapping contributions from the prolate and the oblate twinned units, and completed with possible dynamical scattering (shown tentatively by much weaker reflections). Two of five prolate (red) and oblate (blue) unit cells are indexed and their twinning plane (tp) indicated.}
\end{figure}

Similar twinned constructions were composed with the oblate golden rhombohedra, representing the interstices between the prolate units of the clusters in contact. Contributions from both polyhedral types to the DPs along the same directions were overlapped and completed with possible dynamical scattering. As an example, the resulting overlapped DP along one of the twelve equivalent five-fold $\hkl<01\phi>$ (i.e. approximately $\hkl<0,34,55>$) QC zone-axes is shown in Fig.~\ref{fig:fig1} \cite{online}. 

To simulate the DPs along any of the $\hkl<111>$ zone axes, all twenty twinned prolate rhombohedra, belonging to a RH were rotated by $37.38^{\circ}$, i.e. the angle between the rhombohedral $\hkl<100>$ and $\hkl<111>$ directions, which correspond to the $\hkl<01\phi>$ five-fold and the $\hkl<111>$ three-fold directions of the RH. By following the same procedure as in case of the five-fold zone axes, DPs along the three-fold axes, or along any other direction can be completed. The calculated DPs along the $\hkl<01\phi>$ and $\hkl<111>$ icosahedral zone axes fit very well with the published experimental DPs of icosahedral QCs \cite{8,22,23,24}. Since none of the published experimental DPs was  indexed, comparisons with less symmetric zones were not performed.

\textit{The translational symmetry and the Fourier transform of an extended structural model.}--Next, the ordering between adjacent clusters, formed around particular RHs, was considered. The apparently lost translational symmetry in QCs was the main reason for classifying them as exceptions with regard to normal single crystals. The distances between multiply twinned five-fold rotational centers vary throughout the QC structure and form clusters of variable sizes around individual RH centers. It was recently shown how the icosahedral QCs grow, how this growth is influenced by defects and how a wrong initial stacking is modified during further growth \cite{29}. Regardless of the actual growing mechanism and the corresponding accommodation of the units composing the structure, the prolate as well as the oblate rhombohedra in contact across adjacent clusters are kept in exact phase relationships. As a result series of distinct periodicities are formed along equivalent directions, interrelated as members of a Fibonacci series. The actual periodicities depend on the orientations of the polyhedra, joining the clusters. The entire crystalline space is thus composed of intergrown twins without any obvious boundaries. The long-range translational order along all equivalent directions depends on the actual sequences of the differently oriented prolate and oblate rhombohedra. Consequently, the reflections in the reciprocal space, belonging to the series of periodicities in the crystal, are sharp and comparable to those of perfectly ordered single crystals. Thus, the apparently lost long-range order, being a result of the five-fold twinning, is replaced by a series of long-periodicities, all of them in proper phase-relationships, like the structure was continuous and not intergrown. 

\begin{figure}[ht!]
\includegraphics[width=7.5cm]{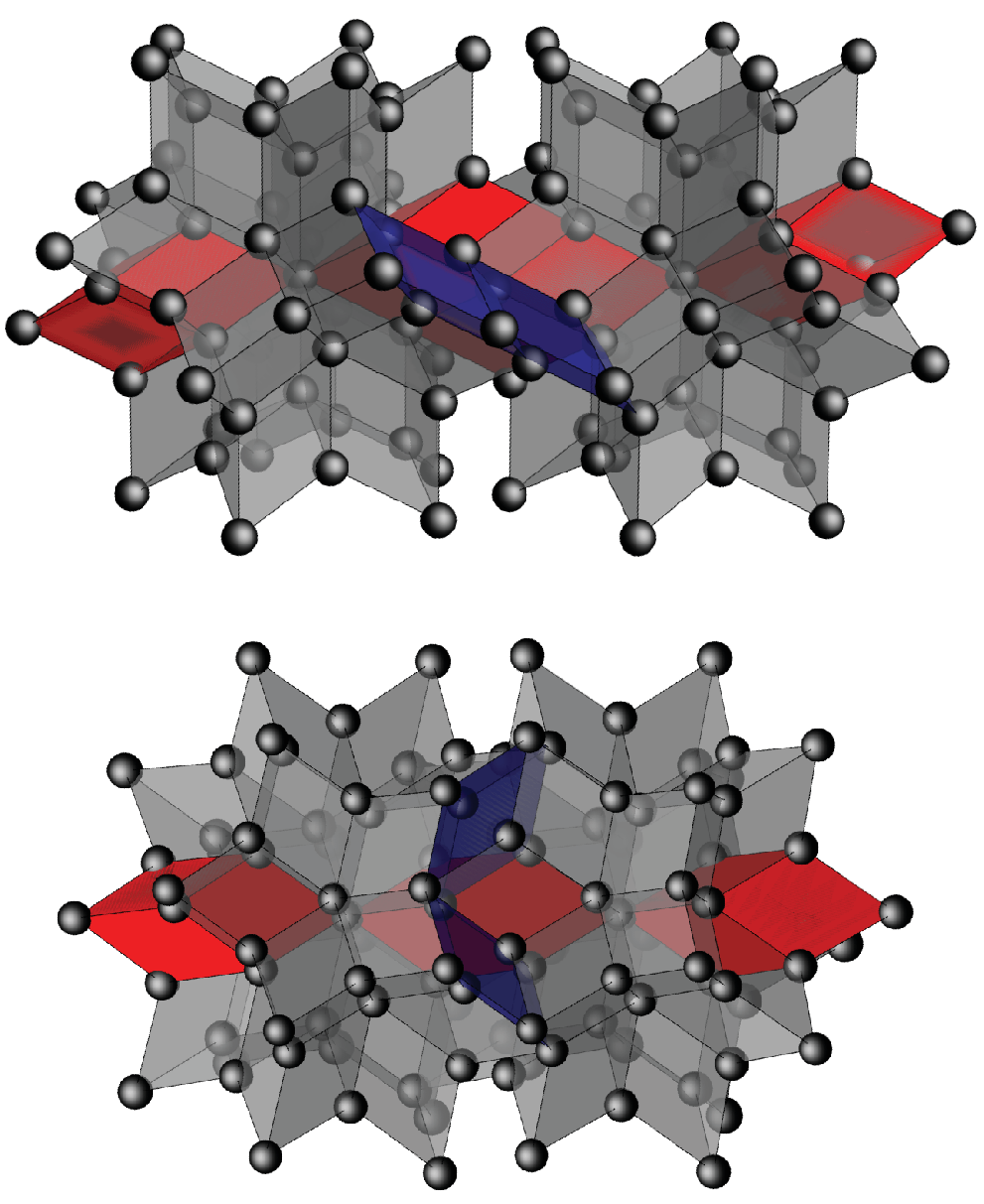}
\caption{\label{fig:fig2} Two possible stackings of adjacent rhombic hexecontahedra. Two of the twenty prolate rhombohedra in both star polyhedra, interrelated by a center of symmetry, are drawn red to point out their stacking. Two oblate rhombohedra are drawn blue.}
\end{figure}

To study the contribution of the specific long-range ordering in the icosahedral QCs, a larger model structure, composed of twinned golden rhombohedra of both types, was constructed of altogether 1272 atoms of a single kind. These atoms represent e.g. in case of MnAl$_{6}$ the disordered constituent Mn and Al atoms in their proper stoichiometric ratio. The structure is shown in Fig.~\ref{fig:fig2}, clearly showing the RH centers and the oblate rhombohedra interconnecting the clusters.  

\begin{figure}[ht!]
\includegraphics[width=7.5cm]{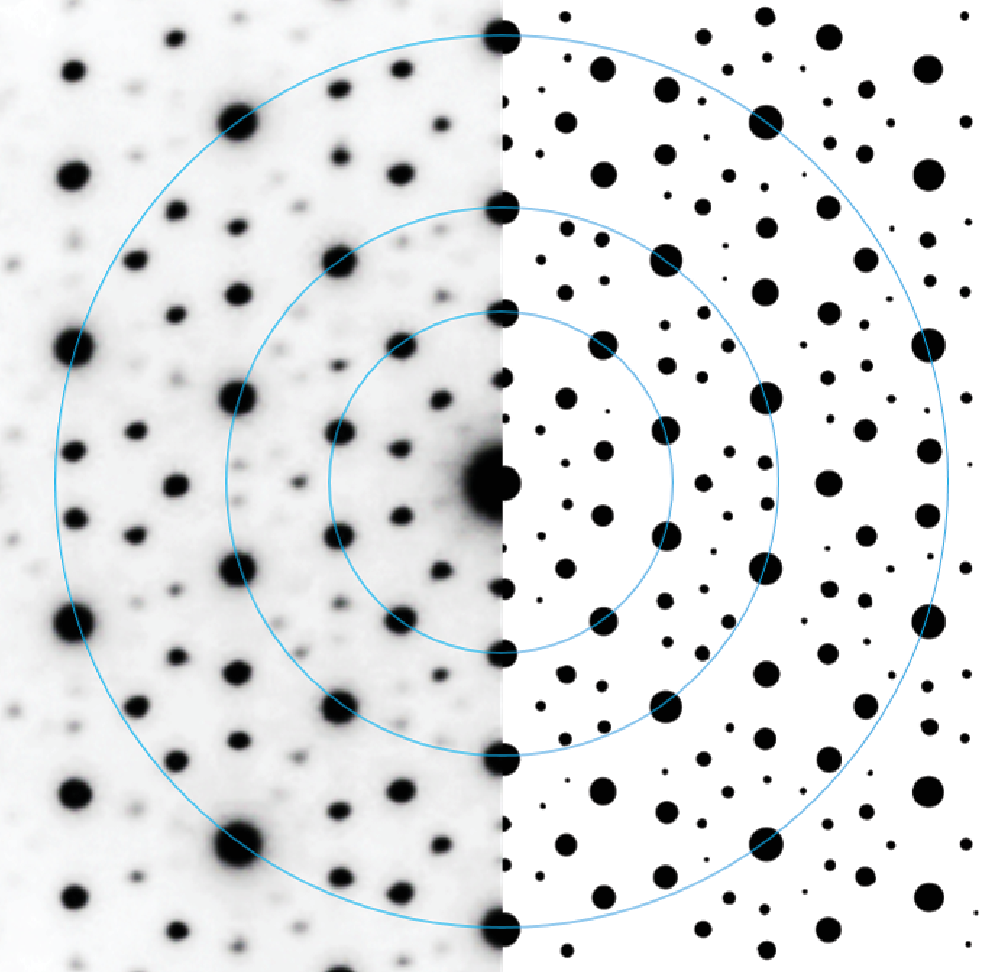}
\caption{\label{fig:fig3} A section through the Fourier transform of the model icosahedral QC structure (right) and the corresponding electron diffraction pattern, recorded along one of the twelve equivalent $\hkl<01\phi>$  icosahedral zone axes (left) \cite{diffraction}.}
\end{figure}

Sections through the three-dimensional Fourier transform (FT) of this model structure, performed perpendicular to one of the twelve equivalent five-fold axes is shown in Fig.~\ref{fig:fig3}. These sections correspond to the experimental DPs along the same zone axis.

\textit{Discussion.}--Five-fold twinning is a known crystal growing mechanism, most often found in nanoparticles and thin films \cite{31}. However, the twinning operation in these cases is applied to a crystal as a whole and thus limited to a single five-fold rotation. The icosahedral QCs with the specific structure can be likewise considered a result of cyclic twinning at the unit-cell level \cite{18a,18b}, i.e. with the twinning operation applied on the primitive golden rhombohedra. If a parent face-centered cubic cell is deformed into four equivalent primitive prolate golden rhombohedra with $\alpha_{p}=\unit[63.43]{^{\circ}}$, the space can be locally filled by forming RHs with twenty twinned prolate rhombohedra each. Clusters, formed around these RH star polyhedra, cannot be stacked together without leaving interstices in the form of oblate golden rhombohedra. As a result the long-range order in icosahedral QCs is in a specific way preserved. Instead of a single periodicity, a series of periodicities, interrelated as members of a Fibonacci series, is formed along equivalent directions. These periodicities depend on the orientation and the sequence of the prolate rhombohedra and the oblate interstices, which fill the remaining space. Consequently, QCs are twinned single crystals with preserved rotational, but in a certain way also translational symmetry. Since the twins are intergrown, they do not show clear twin-boundaries. The related long-range periodicities are determined by combinations of long and short segments, usually considered as sections through a higher-dimensional space\cite{2}. Intergrowth in QCs was suggested before already, e.g. by Mackay \cite{18c} and in one of the papers \cite{18d}, where the QC structure was considered an interpenetrating incommensurately modulated structure, described as a three-dimensional section through a higher-dimensional space \cite{18d,18e,18f,18g}. The related periodicities were described as structural modulations, induced by interacting intergrown subsystems \cite{18d}. However, though mathematically elegant, descriptions in higher dimensions are connected with problems, lacking i.a. exact information on structural parameters, composition and consequently structure-to-properties relationships. A description by means of cyclic unit-cell twinning is fully equivalent with all other approaches and the structure can be considered by means of three-dimensional crystallography with all its advantages. There is also no need for considering the QC structure as being incommensurately modulated. Numerous cases are known, e.g. many quasi one-dimensional  modulated structures \cite{30}, where descriptions in higher dimensions can be replaced by alternative notations, describing separately the basic and the modulated structures. The  icosahedral QCs represents yet another case, where the incommensurate modulation can be replaced by a series of translational periodicities.

Pauling’s approach to the problem was obviously not adequate. Regardless of all arguments brought up in the debates that followed the discovery of QCs, he had to run into problems, because he tried to prove his ideas by applying twinning on huge icosahedral unit cells. There are no forces in nature, able to act between atoms at distances of the order of nanometers and larger. Structures with huge unit cells can only be formed as a result of two or more competing mechanisms, whose relatively short periodicities coincide at large distances. That is also what makes the QC structure a specific case.

Multiple unit-cell twinning must be energetically favorable in comparison with the untwinned parent structure. Whether a parent structure will collapse into a less symmetric one will in metallic alloys depend on the size of the constituent atoms and the stoichiometry of the compound. Twinning of golden rhombohedra is thus a result of the need to reduce the energy and to fill the space accordingly.

It is well known that the DPs of high-quality QCs show exceptionally sharp reflections, whose full widths at half maxima of less than $0.001^{\circ}$ are of the order of the x-ray instrumental resolution \cite{26,27}. Both, the sharp reflections and the very clear background are proofs of a near perfect ordering over large distances, comparable to the best ordered metallic crystals. This is supported by the fact that any multiple scattering coincides with the reflections belonging to distinct long-range periodicities of the Fibonacci series.

\textit{Conclusions.}--It is shown that tiling in icosahedral QCs is fully equivalent to cyclic twinning at the unit-cell level, with the twinning operation applied on primitive prolate golden rhombohedra. Multiple twinning of these rhombohedra forms centers with five-fold rotational symmetry and results in distinct long-range periodicities, interrelated as members of a Fibonacci series. 
The space is locally filled with RH stars of twenty prolate golden rhombohedra, while the interstices between adjacent clusters, formed around the RH centers, represent oblate golden rhombohedra of thirty possible orientations.
The intergrown twinned QC structure shows no explicit twin-boundaries and can be properly described by means of the three-dimensional crystallography.\\

\begin{acknowledgments}
We are thankful to dr. G. Dražič for supplying valuable information on electron diffraction from quasicrystals and to the Slovenian Research Agency (ARRS) for its financial support under Program No. P1-0099.
\end{acknowledgments}

\bibliography{references}

\begin{thebibliography}{39}%
\makeatletter
\providecommand \@ifxundefined [1]{%
 \@ifx{#1\undefined}
}%
\providecommand \@ifnum [1]{%
 \ifnum #1\expandafter \@firstoftwo
 \else \expandafter \@secondoftwo
 \fi
}%
\providecommand \@ifx [1]{%
 \ifx #1\expandafter \@firstoftwo
 \else \expandafter \@secondoftwo
 \fi
}%
\providecommand \natexlab [1]{#1}%
\providecommand \enquote  [1]{``#1''}%
\providecommand \bibnamefont  [1]{#1}%
\providecommand \bibfnamefont [1]{#1}%
\providecommand \citenamefont [1]{#1}%
\providecommand \href@noop [0]{\@secondoftwo}%
\providecommand \href [0]{\begingroup \@sanitize@url \@href}%
\providecommand \@href[1]{\@@startlink{#1}\@@href}%
\providecommand \@@href[1]{\endgroup#1\@@endlink}%
\providecommand \@sanitize@url [0]{\catcode `\\12\catcode `\$12\catcode
  `\&12\catcode `\#12\catcode `\^12\catcode `\_12\catcode `\%12\relax}%
\providecommand \@@startlink[1]{}%
\providecommand \@@endlink[0]{}%
\providecommand \url  [0]{\begingroup\@sanitize@url \@url }%
\providecommand \@url [1]{\endgroup\@href {#1}{\urlprefix }}%
\providecommand \urlprefix  [0]{URL }%
\providecommand \Eprint [0]{\href }%
\providecommand \doibase [0]{http://dx.doi.org/}%
\providecommand \selectlanguage [0]{\@gobble}%
\providecommand \bibinfo  [0]{\@secondoftwo}%
\providecommand \bibfield  [0]{\@secondoftwo}%
\providecommand \translation [1]{[#1]}%
\providecommand \BibitemOpen [0]{}%
\providecommand \bibitemStop [0]{}%
\providecommand \bibitemNoStop [0]{.\EOS\space}%
\providecommand \EOS [0]{\spacefactor3000\relax}%
\providecommand \BibitemShut  [1]{\csname bibitem#1\endcsname}%
\let\auto@bib@innerbib\@empty
\bibitem [{\citenamefont {Shechtman}\ \emph {et~al.}(1984)\citenamefont
  {Shechtman}, \citenamefont {Blech}, \citenamefont {Gratias},\ and\
  \citenamefont {Cahn}}]{1}%
  \BibitemOpen
  \bibfield  {author} {\bibinfo {author} {\bibfnamefont {D.}~\bibnamefont
  {Shechtman}}, \bibinfo {author} {\bibfnamefont {I.}~\bibnamefont {Blech}},
  \bibinfo {author} {\bibfnamefont {D.}~\bibnamefont {Gratias}}, \ and\
  \bibinfo {author} {\bibfnamefont {J.~W.}\ \bibnamefont {Cahn}},\ }\href
  {\doibase 10.1103/PhysRevLett.53.1951} {\bibfield  {journal} {\bibinfo
  {journal} {Phys. Rev. Lett.}\ }\textbf {\bibinfo {volume} {53}},\ \bibinfo
  {pages} {1951} (\bibinfo {year} {1984})}\BibitemShut {NoStop}%
\bibitem [{\citenamefont {Gratias}(2012)}]{2}%
  \BibitemOpen
  \bibfield  {author} {\bibinfo {author} {\bibfnamefont {D.}~\bibnamefont
  {Gratias}},\ }\href {\doibase 10.1051/epn/2012502} {\bibfield  {journal}
  {\bibinfo  {journal} {Europhysics News}\ }\textbf {\bibinfo {volume} {43}},\
  \bibinfo {pages} {26} (\bibinfo {year} {2012})}\BibitemShut {NoStop}%
\bibitem [{3()}]{3}%
  \BibitemOpen
  \href {https://paulingblog.wordpress.com/tag/quasicrystals/ Part 1:
  2012/05/02 The Quasicrystals Puzzle: An Introduction; Part 2: 2012/05/09 The
  Pauling Theory of Quasicrystals; Part 3: 2012/05/16 Shechtman and Pauling
  Debate Quasicrystal Thory; Part 4: 2012/05/23/ Closing the Book on
  Quasicrystals} {\enquote {\bibinfo {title} {The {Pauling} {Blog}, {Part
  1-4}},}\ }\bibinfo {howpublished}
  {\url{https://paulingblog.wordpress.com/tag/quasicrystals}}\BibitemShut
  {NoStop}%
\bibitem [{\citenamefont {Kramer}(2010)}]{3a}%
  \BibitemOpen
  \bibfield  {author} {\bibinfo {author} {\bibfnamefont {P.}~\bibnamefont
  {Kramer}},\ }\href@noop {} {\bibfield  {journal} {\bibinfo  {journal}
  {arXiv:1101v1}\ } (\bibinfo {year} {2010})}\BibitemShut {NoStop}%
\bibitem [{\citenamefont {Shechtman}\ and\ \citenamefont {Blech}(1985)}]{4}%
  \BibitemOpen
  \bibfield  {author} {\bibinfo {author} {\bibfnamefont {D.}~\bibnamefont
  {Shechtman}}\ and\ \bibinfo {author} {\bibfnamefont {I.~A.}\ \bibnamefont
  {Blech}},\ }\href {\doibase 10.1007/BF02811670} {\bibfield  {journal}
  {\bibinfo  {journal} {Metallurgical Transactions A}\ }\textbf {\bibinfo
  {volume} {16}},\ \bibinfo {pages} {1005} (\bibinfo {year}
  {1985})}\BibitemShut {NoStop}%
\bibitem [{\citenamefont {Stephens}\ and\ \citenamefont {Goldman}(1986)}]{5}%
  \BibitemOpen
  \bibfield  {author} {\bibinfo {author} {\bibfnamefont {P.~W.}\ \bibnamefont
  {Stephens}}\ and\ \bibinfo {author} {\bibfnamefont {A.~I.}\ \bibnamefont
  {Goldman}},\ }\href {\doibase 10.1103/PhysRevLett.56.1168} {\bibfield
  {journal} {\bibinfo  {journal} {Phys. Rev. Lett.}\ }\textbf {\bibinfo
  {volume} {56}},\ \bibinfo {pages} {1168} (\bibinfo {year}
  {1986})}\BibitemShut {NoStop}%
\bibitem [{\citenamefont {Pauling}(1985)}]{6}%
  \BibitemOpen
  \bibfield  {author} {\bibinfo {author} {\bibfnamefont {L.}~\bibnamefont
  {Pauling}},\ }\href {http://dx.doi.org/10.1038/317512a0} {\bibfield
  {journal} {\bibinfo  {journal} {Nature}\ }\textbf {\bibinfo {volume} {317}},\
  \bibinfo {pages} {512} (\bibinfo {year} {1985})}\BibitemShut {NoStop}%
\bibitem [{\citenamefont {Pauling}(1986)}]{7}%
  \BibitemOpen
  \bibfield  {author} {\bibinfo {author} {\bibfnamefont {L.}~\bibnamefont
  {Pauling}},\ }\href@noop {} {\bibfield  {journal} {\bibinfo  {journal}
  {Science News}\ }\textbf {\bibinfo {volume} {129}} (\bibinfo {year}
  {1986})}\BibitemShut {NoStop}%
\bibitem [{\citenamefont {Pauling}(1987)}]{8}%
  \BibitemOpen
  \bibfield  {author} {\bibinfo {author} {\bibfnamefont {L.}~\bibnamefont
  {Pauling}},\ }\href {\doibase 10.1103/PhysRevLett.58.365} {\bibfield
  {journal} {\bibinfo  {journal} {Phys. Rev. Lett.}\ }\textbf {\bibinfo
  {volume} {58}},\ \bibinfo {pages} {365} (\bibinfo {year} {1987})}\BibitemShut
  {NoStop}%
\bibitem [{\citenamefont {Pauling}(1989)}]{9}%
  \BibitemOpen
  \bibfield  {author} {\bibinfo {author} {\bibfnamefont {L.}~\bibnamefont
  {Pauling}},\ }\href@noop {} {\bibfield  {journal} {\bibinfo  {journal} {Proc.
  Nat. Acad. Sci. USA}\ }\textbf {\bibinfo {volume} {86}},\ \bibinfo {pages}
  {8595} (\bibinfo {year} {1989})}\BibitemShut {NoStop}%
\bibitem [{\citenamefont {Steinhardt}\ and\ \citenamefont
  {Ostlund}(1987)}]{16}%
  \BibitemOpen
  \bibinfo {editor} {\bibfnamefont {P.}~\bibnamefont {Steinhardt}}\ and\
  \bibinfo {editor} {\bibfnamefont {S.}~\bibnamefont {Ostlund}},\ eds.,\
  \href@noop {} {\emph {\bibinfo {title} {The Physics of Quasicrystals}}}\
  (\bibinfo  {publisher} {World Scientific Publishing},\ \bibinfo {year}
  {1987})\ \bibinfo {note} {310-12}\BibitemShut {NoStop}%
\bibitem [{\citenamefont {Cahn}\ \emph {et~al.}(1986)\citenamefont {Cahn},
  \citenamefont {Gratias},\ and\ \citenamefont {Shechtman}}]{10}%
  \BibitemOpen
  \bibfield  {author} {\bibinfo {author} {\bibfnamefont {J.}~\bibnamefont
  {Cahn}}, \bibinfo {author} {\bibfnamefont {D.}~\bibnamefont {Gratias}}, \
  and\ \bibinfo {author} {\bibfnamefont {D.}~\bibnamefont {Shechtman}},\ }\href
  {http://dx.doi.org/10.1038/319102a0} {\bibfield  {journal} {\bibinfo
  {journal} {Nature}\ }\textbf {\bibinfo {volume} {319}},\ \bibinfo {pages}
  {102} (\bibinfo {year} {1986})}\BibitemShut {NoStop}%
\bibitem [{\citenamefont {Mackay}(1986)}]{11}%
  \BibitemOpen
  \bibfield  {author} {\bibinfo {author} {\bibfnamefont {A.~L.}\ \bibnamefont
  {Mackay}},\ }\href {\doibase doi:10.1038/319103a0} {\bibfield  {journal}
  {\bibinfo  {journal} {Nature}\ }\textbf {\bibinfo {volume} {319}},\ \bibinfo
  {pages} {103} (\bibinfo {year} {1986})}\BibitemShut {NoStop}%
\bibitem [{\citenamefont {Bancel}\ \emph {et~al.}(1986)\citenamefont {Bancel},
  \citenamefont {Heiney}, \citenamefont {Stephens},\ and\ \citenamefont
  {Goldman}}]{12}%
  \BibitemOpen
  \bibfield  {author} {\bibinfo {author} {\bibfnamefont {P.~A.}\ \bibnamefont
  {Bancel}}, \bibinfo {author} {\bibfnamefont {P.~A.}\ \bibnamefont {Heiney}},
  \bibinfo {author} {\bibfnamefont {P.~W.}\ \bibnamefont {Stephens}}, \ and\
  \bibinfo {author} {\bibfnamefont {A.~I.}\ \bibnamefont {Goldman}},\
  }\href@noop {} {\bibfield  {journal} {\bibinfo  {journal} {Nature}\ }\textbf
  {\bibinfo {volume} {319}},\ \bibinfo {pages} {103} (\bibinfo {year}
  {1986})}\BibitemShut {NoStop}%
\bibitem [{\citenamefont {Berezin}(1986)}]{13}%
  \BibitemOpen
  \bibfield  {author} {\bibinfo {author} {\bibfnamefont {A.~A.}\ \bibnamefont
  {Berezin}},\ }\href@noop {} {\bibfield  {journal} {\bibinfo  {journal}
  {Nature}\ }\textbf {\bibinfo {volume} {319}},\ \bibinfo {pages} {104}
  (\bibinfo {year} {1986})}\BibitemShut {NoStop}%
\bibitem [{\citenamefont {Heiney}\ \emph {et~al.}(1987)\citenamefont {Heiney},
  \citenamefont {Bancel},\ and\ \citenamefont {Horn}}]{14}%
  \BibitemOpen
  \bibfield  {author} {\bibinfo {author} {\bibfnamefont {P.~A.}\ \bibnamefont
  {Heiney}}, \bibinfo {author} {\bibfnamefont {P.~A.}\ \bibnamefont {Bancel}},
  \ and\ \bibinfo {author} {\bibfnamefont {P.~M.}\ \bibnamefont {Horn}},\
  }\href {\doibase 10.1103/PhysRevLett.59.2119} {\bibfield  {journal} {\bibinfo
   {journal} {Phys. Rev. Lett.}\ }\textbf {\bibinfo {volume} {59}},\ \bibinfo
  {pages} {2119} (\bibinfo {year} {1987})}\BibitemShut {NoStop}%
\bibitem [{\citenamefont {Bancel}\ \emph {et~al.}(1989)\citenamefont {Bancel},
  \citenamefont {Heiney}, \citenamefont {Horn},\ and\ \citenamefont
  {Steinhardt}}]{15}%
  \BibitemOpen
  \bibfield  {author} {\bibinfo {author} {\bibfnamefont {P.~A.}\ \bibnamefont
  {Bancel}}, \bibinfo {author} {\bibfnamefont {P.~A.}\ \bibnamefont {Heiney}},
  \bibinfo {author} {\bibfnamefont {P.~M.}\ \bibnamefont {Horn}}, \ and\
  \bibinfo {author} {\bibfnamefont {P.~J.}\ \bibnamefont {Steinhardt}},\
  }\href@noop {} {\bibfield  {journal} {\bibinfo  {journal} {Proc. Nat. Acad.
  Sci. USA}\ }\textbf {\bibinfo {volume} {86(22)}},\ \bibinfo {pages} {8600}
  (\bibinfo {year} {1989})}\BibitemShut {NoStop}%
\bibitem [{\citenamefont {Lidin}\ \emph {et~al.}(1989)\citenamefont {Lidin},
  \citenamefont {Andersson}, \citenamefont {Bovin}, \citenamefont {Malm},\ and\
  \citenamefont {Terasaki}}]{15a}%
  \BibitemOpen
  \bibfield  {author} {\bibinfo {author} {\bibfnamefont {S.}~\bibnamefont
  {Lidin}}, \bibinfo {author} {\bibfnamefont {S.}~\bibnamefont {Andersson}},
  \bibinfo {author} {\bibfnamefont {J.-O.}\ \bibnamefont {Bovin}}, \bibinfo
  {author} {\bibfnamefont {J.~O.}\ \bibnamefont {Malm}}, \ and\ \bibinfo
  {author} {\bibfnamefont {O.}~\bibnamefont {Terasaki}},\ }\href {\doibase
  10.1107/S0108767389012018} {\bibfield  {journal} {\bibinfo  {journal} {Acta
  Crystallographica Section A}\ }\textbf {\bibinfo {volume} {45}},\ \bibinfo
  {pages} {fc33} (\bibinfo {year} {1989})}\BibitemShut {NoStop}%
\bibitem [{\citenamefont {Lord}\ \emph {et~al.}(2000)\citenamefont {Lord},
  \citenamefont {Ranganathan},\ and\ \citenamefont {Kulkarni}}]{17}%
  \BibitemOpen
  \bibfield  {author} {\bibinfo {author} {\bibfnamefont {E.~A.}\ \bibnamefont
  {Lord}}, \bibinfo {author} {\bibfnamefont {S.}~\bibnamefont {Ranganathan}}, \
  and\ \bibinfo {author} {\bibfnamefont {U.~D.}\ \bibnamefont {Kulkarni}},\
  }\href@noop {} {\bibfield  {journal} {\bibinfo  {journal} {Current Science}\
  }\textbf {\bibinfo {volume} {78}},\ \bibinfo {pages} {64} (\bibinfo {year}
  {2000})}\BibitemShut {NoStop}%
\bibitem [{\citenamefont {Andersson}\ and\ \citenamefont {Hyde}(1974)}]{18a}%
  \BibitemOpen
  \bibfield  {author} {\bibinfo {author} {\bibfnamefont {S.}~\bibnamefont
  {Andersson}}\ and\ \bibinfo {author} {\bibfnamefont {B.}~\bibnamefont
  {Hyde}},\ }\href {\doibase http://dx.doi.org/10.1016/0022-4596(74)90059-0}
  {\bibfield  {journal} {\bibinfo  {journal} {Journal of Solid State
  Chemistry}\ }\textbf {\bibinfo {volume} {9}},\ \bibinfo {pages} {92 }
  (\bibinfo {year} {1974})}\BibitemShut {NoStop}%
\bibitem [{\citenamefont {Hyde}\ and\ \citenamefont {Andersson}(1989)}]{18b}%
  \BibitemOpen
  \bibfield  {author} {\bibinfo {author} {\bibfnamefont {B.~G.}\ \bibnamefont
  {Hyde}}\ and\ \bibinfo {author} {\bibfnamefont {S.}~\bibnamefont
  {Andersson}},\ }\href@noop {} {\emph {\bibinfo {title} {Inorganic Crystal
  Structures}}}\ (\bibinfo  {publisher} {John Wiley \& Sons, N.Y.},\ \bibinfo
  {year} {1989})\BibitemShut {NoStop}%
\bibitem [{\citenamefont {Guyot}(1987)}]{28}%
  \BibitemOpen
  \bibfield  {author} {\bibinfo {author} {\bibfnamefont {P.}~\bibnamefont
  {Guyot}},\ }\href {\doibase 10.1038/326640a0} {\bibfield  {journal} {\bibinfo
   {journal} {Nature}\ ,\ \bibinfo {pages} {640}} (\bibinfo {year}
  {1987})}\BibitemShut {NoStop}%
\bibitem [{\citenamefont {Weber}(2015)}]{20}%
  \BibitemOpen
  \bibfield  {author} {\bibinfo {author} {\bibfnamefont {S.}~\bibnamefont
  {Weber}},\ }\href@noop {} {\enquote {\bibinfo {title} {Quasicrystals},}\
  }\bibinfo {howpublished} {\url{http://www.jcrystal.com/steffenweber/qc.html}}
  (\bibinfo {year} {2015})\BibitemShut {NoStop}%
\bibitem [{\citenamefont {Knott}(2009)}]{21}%
  \BibitemOpen
  \bibfield  {author} {\bibinfo {author} {\bibfnamefont {R.}~\bibnamefont
  {Knott}},\ }\href
  {http://www.maths.surrey.ac.uk/hosted-sites/R.Knott/Fibonacci/phi3DGeom.html#phi3D}
  {\enquote {\bibinfo {title} {The golden geometry of solids or phi in 3
  dimensions},}\ }\bibinfo {howpublished}
  {\url{http://www.maths.surrey.ac.uk/hosted-sites/R.Knott/Fibonacci/phi3DGeom.html#phi3D}}
  (\bibinfo {year} {2009})\BibitemShut {NoStop}%
\bibitem [{onl()}]{online}%
  \BibitemOpen
  \href@noop {} {}\bibinfo {howpublished} {See supplementary material at [URL
  will be inserted by publisher] for details.}\BibitemShut {Stop}%
\bibitem [{\citenamefont {Bindi}\ \emph {et~al.}(2009)\citenamefont {Bindi},
  \citenamefont {Steinhardt}, \citenamefont {Yao},\ and\ \citenamefont
  {Lu}}]{22}%
  \BibitemOpen
  \bibfield  {author} {\bibinfo {author} {\bibfnamefont {L.}~\bibnamefont
  {Bindi}}, \bibinfo {author} {\bibfnamefont {P.~J.}\ \bibnamefont
  {Steinhardt}}, \bibinfo {author} {\bibfnamefont {N.}~\bibnamefont {Yao}}, \
  and\ \bibinfo {author} {\bibfnamefont {P.~J.}\ \bibnamefont {Lu}},\ }\href
  {\doibase 10.1126/science.1170827} {\bibfield  {journal} {\bibinfo  {journal}
  {Science}\ }\textbf {\bibinfo {volume} {324}},\ \bibinfo {pages} {1306}
  (\bibinfo {year} {2009})}\BibitemShut {NoStop}%
\bibitem [{\citenamefont {Guo}\ \emph {et~al.}(2000)\citenamefont {Guo},
  \citenamefont {Abe},\ and\ \citenamefont {Tsai}}]{23}%
  \BibitemOpen
  \bibfield  {author} {\bibinfo {author} {\bibfnamefont {J.~Q.}\ \bibnamefont
  {Guo}}, \bibinfo {author} {\bibfnamefont {E.}~\bibnamefont {Abe}}, \ and\
  \bibinfo {author} {\bibfnamefont {A.~P.}\ \bibnamefont {Tsai}},\ }\href
  {\doibase 10.1103/PhysRevB.62.R14605} {\bibfield  {journal} {\bibinfo
  {journal} {Phys. Rev. B}\ }\textbf {\bibinfo {volume} {62}},\ \bibinfo
  {pages} {R14605} (\bibinfo {year} {2000})}\BibitemShut {NoStop}%
\bibitem [{24()}]{24}%
  \BibitemOpen
  \href@noop {} {\enquote {\bibinfo {title} {Quasicrystal - {Wikipedia}},}\
  }\bibinfo {howpublished}
  {\url{http://en.wikipedia.org/wiki/Quasicrystal}}\BibitemShut {NoStop}%
\bibitem [{\citenamefont {Hann}\ \emph {et~al.}(2016)\citenamefont {Hann},
  \citenamefont {Socolar},\ and\ \citenamefont {Steinhardt}}]{29}%
  \BibitemOpen
  \bibfield  {author} {\bibinfo {author} {\bibfnamefont {C.~T.}\ \bibnamefont
  {Hann}}, \bibinfo {author} {\bibfnamefont {J.~E.~S.}\ \bibnamefont
  {Socolar}}, \ and\ \bibinfo {author} {\bibfnamefont {P.~J.}\ \bibnamefont
  {Steinhardt}},\ }\href {\doibase 10.1103/PhysRevB.94.014113} {\bibfield
  {journal} {\bibinfo  {journal} {Phys. Rev. B}\ }\textbf {\bibinfo {volume}
  {94}},\ \bibinfo {pages} {014113} (\bibinfo {year} {2016})}\BibitemShut
  {NoStop}%
\bibitem [{dif()}]{diffraction}%
  \BibitemOpen
  \href@noop {} {\enquote {\bibinfo {title} {Electron diffraction pattern of an
  icosahedral zn-mg-ho quasicrystal, cc by-nc-sa 2.0},}\ }\bibinfo
  {howpublished}
  {\url{https://commons.wikimedia.org/wiki/File:Zn-Mg-HoDiffraction.JPG}}\BibitemShut
  {NoStop}%
\bibitem [{\citenamefont {Hofmeister}(2004)}]{31}%
  \BibitemOpen
  \bibfield  {author} {\bibinfo {author} {\bibfnamefont {H.}~\bibnamefont
  {Hofmeister}},\ }\href@noop {} {\emph {\bibinfo {title} {Encyclopedia of
  Nanoscience and nanotechnology vol. 3}}}\ (\bibinfo  {publisher} {American
  Scientific Publishers, Stevenson Ranch},\ \bibinfo {year} {2004})\ pp.\
  \bibinfo {pages} {431--452}\BibitemShut {NoStop}%
\bibitem [{\citenamefont {Mackay}(1987)}]{18c}%
  \BibitemOpen
  \bibfield  {author} {\bibinfo {author} {\bibfnamefont {A.~L.}\ \bibnamefont
  {Mackay}},\ }\href {\doibase 10.1111/j.1365-2818.1987.tb01347.x} {\bibfield
  {journal} {\bibinfo  {journal} {Journal of Microscopy}\ }\textbf {\bibinfo
  {volume} {146}},\ \bibinfo {pages} {233} (\bibinfo {year}
  {1987})}\BibitemShut {NoStop}%
\bibitem [{\citenamefont {Spal}(1986)}]{18d}%
  \BibitemOpen
  \bibfield  {author} {\bibinfo {author} {\bibfnamefont {R.~D.}\ \bibnamefont
  {Spal}},\ }\href {\doibase 10.1103/PhysRevLett.56.1823} {\bibfield  {journal}
  {\bibinfo  {journal} {Phys. Rev. Lett.}\ }\textbf {\bibinfo {volume} {56}},\
  \bibinfo {pages} {1823} (\bibinfo {year} {1986})}\BibitemShut {NoStop}%
\bibitem [{\citenamefont {{CAHN, J. W.}}\ and\ \citenamefont {{GRATIAS,
  D.}}(1986)}]{18e}%
  \BibitemOpen
  \bibfield  {author} {\bibinfo {author} {\bibnamefont {{CAHN, J. W.}}}\ and\
  \bibinfo {author} {\bibnamefont {{GRATIAS, D.}}},\ }\href {\doibase
  10.1051/jphyscol:1986342} {\bibfield  {journal} {\bibinfo  {journal} {J.
  Phys. Colloques}\ }\textbf {\bibinfo {volume} {47}},\ \bibinfo {pages} {C3}
  (\bibinfo {year} {1986})}\BibitemShut {NoStop}%
\bibitem [{\citenamefont {{Cahn, J.W.}}\ \emph {et~al.}(1988)\citenamefont
  {{Cahn, J.W.}}, \citenamefont {{Gratias, D.}},\ and\ \citenamefont {{Mozer,
  B.}}}]{18f}%
  \BibitemOpen
  \bibfield  {author} {\bibinfo {author} {\bibnamefont {{Cahn, J.W.}}},
  \bibinfo {author} {\bibnamefont {{Gratias, D.}}}, \ and\ \bibinfo {author}
  {\bibnamefont {{Mozer, B.}}},\ }\href {\doibase
  10.1051/jphys:019880049070122500} {\bibfield  {journal} {\bibinfo  {journal}
  {J. Phys. France}\ }\textbf {\bibinfo {volume} {49}},\ \bibinfo {pages}
  {1225} (\bibinfo {year} {1988})}\BibitemShut {NoStop}%
\bibitem [{\citenamefont {{Duneau, M.}}\ and\ \citenamefont {{Oguey,
  C.}}(1989)}]{18g}%
  \BibitemOpen
  \bibfield  {author} {\bibinfo {author} {\bibnamefont {{Duneau, M.}}}\ and\
  \bibinfo {author} {\bibnamefont {{Oguey, C.}}},\ }\href {\doibase
  10.1051/jphys:01989005002013500} {\bibfield  {journal} {\bibinfo  {journal}
  {J. Phys. France}\ }\textbf {\bibinfo {volume} {50}},\ \bibinfo {pages} {135}
  (\bibinfo {year} {1989})}\BibitemShut {NoStop}%
\bibitem [{\citenamefont {Monceau}(2012)}]{30}%
  \BibitemOpen
  \bibfield  {author} {\bibinfo {author} {\bibfnamefont {P.}~\bibnamefont
  {Monceau}},\ }\href {\doibase 10.1080/00018732.2012.719674} {\bibfield
  {journal} {\bibinfo  {journal} {Advances in Physics}\ }\textbf {\bibinfo
  {volume} {61}},\ \bibinfo {pages} {325} (\bibinfo {year} {2012})},\ \Eprint
  {http://arxiv.org/abs/http://dx.doi.org/10.1080/00018732.2012.719674}
  {http://dx.doi.org/10.1080/00018732.2012.719674} \BibitemShut {NoStop}%
\bibitem [{\citenamefont {Abe}\ \emph {et~al.}(2004)\citenamefont {Abe},
  \citenamefont {Yanfa},\ and\ \citenamefont {J.~Pennycook}}]{26}%
  \BibitemOpen
  \bibfield  {author} {\bibinfo {author} {\bibfnamefont {E.}~\bibnamefont
  {Abe}}, \bibinfo {author} {\bibfnamefont {Y.}~\bibnamefont {Yanfa}}, \ and\
  \bibinfo {author} {\bibfnamefont {S.}~\bibnamefont {J.~Pennycook}},\ }\href
  {\doibase 10.1038/nmat1244} {\bibfield  {journal} {\bibinfo  {journal}
  {Nature Materials}\ ,\ \bibinfo {pages} {759}} (\bibinfo {year}
  {2004})}\BibitemShut {NoStop}%
\bibitem [{\citenamefont {Kycia}\ \emph {et~al.}(1993)\citenamefont {Kycia},
  \citenamefont {Goldman}, \citenamefont {Lograsso}, \citenamefont {Delaney},
  \citenamefont {Black}, \citenamefont {Sutton}, \citenamefont {Dufresne},
  \citenamefont {Br\"uning},\ and\ \citenamefont {Rodricks}}]{27}%
  \BibitemOpen
  \bibfield  {author} {\bibinfo {author} {\bibfnamefont {S.~W.}\ \bibnamefont
  {Kycia}}, \bibinfo {author} {\bibfnamefont {A.~I.}\ \bibnamefont {Goldman}},
  \bibinfo {author} {\bibfnamefont {T.~A.}\ \bibnamefont {Lograsso}}, \bibinfo
  {author} {\bibfnamefont {D.~W.}\ \bibnamefont {Delaney}}, \bibinfo {author}
  {\bibfnamefont {D.}~\bibnamefont {Black}}, \bibinfo {author} {\bibfnamefont
  {M.}~\bibnamefont {Sutton}}, \bibinfo {author} {\bibfnamefont
  {E.}~\bibnamefont {Dufresne}}, \bibinfo {author} {\bibfnamefont
  {R.}~\bibnamefont {Br\"uning}}, \ and\ \bibinfo {author} {\bibfnamefont
  {B.}~\bibnamefont {Rodricks}},\ }\href {\doibase 10.1103/PhysRevB.48.3544}
  {\bibfield  {journal} {\bibinfo  {journal} {Phys. Rev. B}\ }\textbf {\bibinfo
  {volume} {48}},\ \bibinfo {pages} {3544} (\bibinfo {year}
  {1993})}\BibitemShut {NoStop}%
\end{thebibliography}%

\end{document}